\documentclass[
               twocolumn,
               noshowpacs,          
               nopreprintnumbers,     
               aps,                 
               prd,                 
               a4paper,             
               superscriptaddress,  
               nofootinbib,         
               tightenlines,        
               floats,floatfix      
               ]{revtex4}

\usepackage{amsmath, amsfonts, amsthm, amssymb, graphicx}

 \usepackage{slashed}

\def\be{\begin{equation}}
\def\ee{\end{equation}}
\def\ba{\begin{eqnarray}}
\def\ea{\end{eqnarray}}

\def\bdm{\begin{displaymath}}
\def\edm{\end{displaymath}}

\def\bq{\begin{quote}}
\def\eq{\end{quote}}

\def\del{\partial}
\def\ltap{\ \raise.3ex\hbox{$<$\kern-.75em\lower1ex\hbox{$\sim$}}\ }
\def\gtap{\ \raise.3ex\hbox{$>$\kern-.75em\lower1ex\hbox{$\sim$}}\ }
\def\gl{\ \raise.5ex\hbox{$>$}\kern-.8em\lower.5ex\hbox{$<$}\ }
\def\roughly#1{\raise.3ex\hbox{$#1$\kern-.75em\lower1ex\hbox{$\sim$}}}

 at 10truept

\newcommand{\beq}{\begin{equation}}
\newcommand{\eeq}{\end{equation}}
\newcommand{\bea}{\begin{eqnarray}}
\newcommand{\eea}{\end{eqnarray}}
\newcommand{\beqa}{\begin{eqnarray}}
\newcommand{\eeqa}{\end{eqnarray}}

\def \del {\partial}

\def \mn {{\mu\nu}}

\begin{document}

\title{Generalised Scale Invariant Theories}

\author{Antonio Padilla}
\email{antonio.padilla@nottingham.ac.uk}
\affiliation{School of Physics and Astronomy, 
University of Nottingham, Nottingham NG7 2RD, UK} 
\author{David Stefanyszyn}
\email{ppxds1@nottingham.ac.uk}
\affiliation{School of Physics and Astronomy, 
University of Nottingham, Nottingham NG7 2RD, UK} 
\author{Minas Tsoukalas}
\email{minasts@cecs.cl} 
\affiliation{Centro de Estudios Cient\' ificos, Casilla 1469, Valdivia, Chile}

\date{\today}

\begin{abstract}
We present the most general actions of a single scalar field and two scalar fields coupled to gravity, consistent with second order field equations in four dimensions, possessing local scale invariance. We apply two different methods to arrive at our results. One method, Ricci gauging, was known to the literature and we find this to produce the same result for the case of one scalar field as a more efficient method presented here. However, we also find our more efficient method to be much more general when we consider two scalar fields. Locally scale invariant actions are also presented for theories with more than  two scalar fields coupled to gravity and we explain how one could construct the most general actions for any number of scalar fields. Our generalised scale invariant actions have obvious applications to early universe cosmology, and include, for example, the Bezrukov-Shaposhnikov action as a subset.
\end{abstract}

\maketitle


\section{Introduction}
It could be said that the Universe is {\it nearly} scale invariant.  This is certainly true of the cosmic microwave background fluctuations recently measured to remarkable accuracy by Planck\cite{Planck}, as well as the Standard Model of Particle Physics, whose classical scale invariance is only spoilt by the Higgs mass.  One might suspect that this is more than a coincidence and that scale invariance has some role to play in our search for a fundamental theory of Nature.  
  
Scale invariant theories have been studied in many different contexts, dating back at least as far as Weyl's attempts to unify gravity with electromagnetism \cite{Weyl}. It has recently been argued by 't Hooft that they could play an important role in understanding black hole phenomena and quantum gravity \cite{thooft}, whilst in early Universe cosmology, there are claims that scale invariance can help us find geodesically complete solutions \cite{Bars1,Bars2} (see, however, \cite{Kallosh} and later \cite{Bars3}). Scale invariance has also been used to identify universality in a wide range of inflationary models \cite{Linde}. However, perhaps the most compelling reason to study scale invariance is within the context of Nature's hierarchies. The Standard Model of Particle Physics and the concordance model in cosmology are both plagued by unnaturally small mass scales, corresponding to the Higgs mass ($\sim 10^{-16} M_{pl}$) and the cosmological constant ($\sim 10^{-60} M_{pl}$) respectively. In a scale invariant theory there are no mass scales and the hope is that when the symmetry is broken such hierarchies emerge naturally (see  eg \cite{def}).  Realising this in practice is  challenging, especially for the latter hierarchy, not least because of the restrictions imposed by Weinberg's famous no-go theorem. Nevertheless, there have been interesting proposals. Recently, one of us developed a model in which the Standard Model vacuum energy is sequestered from gravity, exploiting, in part, global scale invariance in the protected matter sector \cite{seq}.  There have also been attempts to use scale invariance to stabilise the Higgs mass (see eg \cite{Nic}). In a series of papers, Shaposhnikov and collaborators have exploited it to propose a complete cosmological model that attempts to  include Higgs inflation, a stable Higgs mass, and dynamical dark energy (see eg \cite{Shap}). For other uses of scale invariance in cosmology and particle physics see, for example, \cite{foot}\cite{jain}\cite{rajpoot}\cite{singh}.  There are, of course, many more interesting applications of scale invariance in  the literature, not least within the context of the AdS/CFT correspondence \cite{adscft}.  There are even applications within biology \cite{biol} and psychology \cite{psy}.
  
In this paper, we identify a plethora of new (multi)scalar and (multi)scalar-tensor theories exhibiting scale invariance. We consider three separate cases in the following sections, namely, single scalar theories, bi-scalar theories and finally, theories with more than two scalars. Indeed, if we wish to preserve second order field equations\footnote{Second order field equations are desirable in order avoid problems with Ostrogradski ghosts \cite{ostro}. If we regard our theory in the language of effective field theories with a cut off, however, one could imagine quantum corrections generating higher order operators suppressed by the cut off scale. These corrections would not introduce any pathologies since the mass of the corresponding ghost is up at the cut off scale \cite{hawking}\cite{cliff}.}, we can say that the theories presented here exhaust all possibilities, at least in four dimensions for the single scalar and bi-scalar cases. This is because our starting point is Horndeski's most general (2nd order) scalar-tensor theory \cite{Horn}. Generality is lost for more than two scalars, however a large class of theories not yet discussed in the literature are presented by exploiting the multi-scalar Horndeski-like  actions presented in \cite{Vish}.  We begin by identifying the subset of  these theories that possess {\it global} scale invariance. The path to {\it local} scale invariance comes in two forms. The first is to simply identify the subset of (multi)Horndeski theories that happens to possess this symmetry. Gravity is known to play a crucial role in terms of gauging the scale invariant $\phi^4$ theory, so the fact that (multi)Horndeski is a gravitational theory is crucial. The second 
 approach is to gauge the globally scale invariant theory directly, using either Weyl gauging, or Ricci gauging \cite{gauging}.  The latter will generically lead to higher order field equations unless they fall into the subset identified in the first approach. Of course, generically we do not expect these theories to maintain scale invariance at the quantum level without being embedded in some larger conformally invariant theory. Indeed it has recently been argued that scale invariance plus unitarity requires conformal invariance in order to be consistent in four dimensions \cite{luty}.

\section{Single scalar theories} 
To illustrate our methods most clearly, we begin by looking at the case of a single scalar, coupled to gravity. In four dimensions, the most general such theory with second order field equations is given by the Horndeski action \cite{Horn}, which we write in the simpler form presented in \cite{dgsz},
\begin{multline}\label{horny}
S_\text{Horndeski}[\phi, g]=\int d^4 x \sqrt{-g} \left[K(\phi, X)-G_3(\phi, X) {\cal E}_1
\right.
\\ \left.
+G_4 (\phi, X)R+G_{4, X} {\cal E}_2  
+G_5(\phi, X) G_{\mu\nu}\nabla^\mu \nabla^\nu \phi-\frac{G_{5, X}}{6} {\cal E}_3 \right]
 \end{multline}
where $X=-\frac{1}{2} (\nabla \phi)^2$, ${\cal E}_n=n! \nabla_{[\mu_1}\nabla^{\mu_1} \phi\cdots \nabla_{\mu_n]}\nabla^{\mu_n} \phi$ and commas denote differentiation i.e. $G_{4,X}=\frac{\del G_4}{\del X}$.

We start by demanding global scale invariance such that under $g_{\mu\nu} \rightarrow \lambda^{2}g_{\mu\nu}$ and $\phi \rightarrow \phi/\lambda$, the Horndeski action is invariant where $\lambda$ is a constant. For a diffeomorphism invariant theory, this rescaling of the metric is equivalent to a rescaling of co-ordinates. Note that we have assumed that the scalar has scaling dimension -1. This can be guaranteed by a simple field redefintion, and given that our starting point is the most general theory, with general potentials, this is without loss of generality. The resulting action is nothing more than one would expect from dimensional analysis with the assumption that no dimensionful couplings can appear in the arbitrary functions and with the scalar field assumed to have a mass dimension equal to 1. We therefore find that the most general globally scale invariant subset of the Horndeski action is given by 
\begin{multline} \label{glob-sing}
S_\text{global}[\phi, g]=\int d^4 x \sqrt{-g} \left[\phi^4 a_2(Y)-\phi a_3(Y){\cal E}_1
\right.
\\ \left.
+a_4(Y)\phi^2 R+\frac{a_4'(Y)}{\phi^2} {\cal E}_2  
+\frac{a_5(Y)}{\phi} G_{\mu\nu}\nabla^\mu \nabla^\nu \phi-\frac{a_5'(Y) }{6\phi^5} {\cal E}_3 \right]
 \end{multline}
 where we have defined a dimensionless quantity $Y=X/\phi^4$. Note that the familiar $-\frac{1}{2} (\nabla \phi)^2 -\mu \phi^4$ theory is readily obtained by taking $a_2=Y-\mu$. 
 We now consider the question of local scale invariance, by which we mean invariance under $g_\mn \to \lambda^2 g_\mn, ~\phi\to \phi/\lambda$, but now $\lambda=\lambda(x)$ . As stated earlier, there are two possible paths to achieving local scale invariance. The first is to identify the subset of our globally scale invariant action (\ref{glob-sing}) that also exhibits local scale invariance. One could examine this directly and establish conditions on the various functions. However, there exists an argument that allows us to go directly to the answer and which will generalise nicely in the multi-scalar cases to be studied later. We denote our locally scale invariant action by $S_\text{local}[\phi, g]$. This action is unchanged by a scale transformation, $g_\mn \to \lambda^2 g_\mn, ~\phi\to \phi/\lambda$, and so choosing  $\lambda(x)=\phi(x)$, we see that $S_\text{local}[\phi, g]=S_\text{local}[1, \tilde g]$, where $\tilde g_\mn=\phi^2 g_\mn$.  To be a subset of the Horndeski action, we know that $S_\text{local}[1, \tilde g]$  must have second order field equations, but by Lovelock's theorem \cite{Lovelock} in four dimensions, the most general diffeomorphism invariant action with second order field equations built out of $\tilde g_\mn$ is 
\be \label{EH}
 S_\text{local}[1, \tilde g]=\int d^4 x \sqrt{-\tilde g} (a \tilde R+b)
\ee
 where $a, b$ are constants, and  $\tilde R$ is the Ricci scalar built out of the metric $\tilde g_\mn$, with a metric connection.  Using the fact that $\tilde g_\mn=\phi^2 g_\mn$, we conclude, after some integration by parts, that the {\it unique} subset of Horndeski exhibiting local scale invariance is given by the well known action for a conformally coupled scalar field \cite{deser},
   \be \label{local}
 S_\text{local}[\phi, g]=-12 a \int d^4 x \sqrt{- g} \left(-\frac{1}{2} (\nabla \phi)^2-\mu  \phi^4 -\frac{1}{12} \phi^2 R  \right)
 \ee
  where $\mu=\frac{b}{12a}$. We note that this result is equivalent to the general action for one scalar presented in \cite{Bars2} where a maximum of two derivatives are allowed in the action. Here we have proven that terms with greater than two derivatives can play no role if we are to keep second order field equations. It is also worth pointing out that one can always generate a scale invariant theory from an action $S[\tilde g]$, where $\tilde g_{\mu\nu}=\phi^2  g_{\mu\nu}$. This  is done in \cite{Tsouk} in order to generate scale invariant theories with a single scalar in arbitrary dimensions with second order field equations. In this paper, we have {\it proven} that the theories generated in \cite{Tsouk} will be the {\it most general} with second order field equations. Similar techniques were used in \cite{alvarez} to express Unimodular Gravity as a theory symmetric under transverse diffeomorphisms and Weyl transformations.

  The second path to local scale invariance is through a straightforward gauging of the global symmetry. To this end we introduce the Weyl vector $W_\mu$ transforming as $W_\mu \to W_\mu +\nabla_\mu \log \lambda$ and the Weyl covariant derivative, $D_\mu=\del_\mu-dW_\mu$ acting on an object with scaling dimension $d$. We obtain a locally scale invariant action by simply replacing all partial derivatives in (\ref{glob-sing}) with the Weyl covariant derivative, or in other words
\ba \label{repl}
\del_\mu \phi& \to& D_\mu\phi=  \del_\mu \phi +W_\mu \phi \nonumber \\
\nabla_\mu \nabla_\nu \phi &\to&  D_\mn \phi=\nabla_\mu \nabla_\nu \phi +\Omega_\mn (W)\phi  \nonumber
\\ 
&&\qquad +(4W_{(\mu} \delta^\alpha_{\nu)}-g_\mn W^{\alpha}) \left(\del_{\alpha}\phi+\frac{1}{2} W_\alpha \phi\right) \nonumber \\
R_{\mu\nu}{}^{\alpha \beta} &\to& {\cal R}_{\mu\nu}{}^{\alpha \beta} =R_{\mu\nu}{}^{\alpha \beta} +4\delta^{[\alpha}_{[\mu} \Omega_{\nu]}{}^{\beta]}  
\ea
where $\Omega_\mn(W)=\nabla_{(\mu} W_{\nu)}+W_\mu W_\nu-\frac{1}{2} g_\mn W^2$. Note that we have dropped terms of the form $\nabla_{[\mu} W_{\nu]}$ in the above, as such terms are Weyl invariant by themselves. In this sense, our action will be the minimally gauged version of (\ref{glob-sing}) rather than the most general Weyl invariant action involving $\phi, g$ and $W$. Our action is, however, still second order and it would therefore be natural to add a gauge invariant kinetic term for the Weyl vector. In any event, our generalised Weyl action for a single scalar field coupled to gravity is given by
\begin{multline} \label{weyl-sing}
S_\text{weyl}[\phi, g, W]=\int d^4 x \sqrt{-g} \left[\phi^4 a_2({\bar Y})-\phi a_3({\bar Y})\bar  { \cal E}_1
\right.
\\ \left.
+a_4({\bar Y})\phi^2 {\cal R}+\frac{a_4'({\bar Y})}{\phi^2} \bar {\cal E}_2  
+\frac{a_5(\bar Y)}{\phi} 
{\cal G}_{\mu\nu}D^\mn \phi-\frac{a_5'(\bar Y) }{6\phi^5} \bar {\cal E}_3 \right] 
 \end{multline}
 where ${\bar Y}=-(D \phi)^2/2\phi^4$, $\bar {\cal E}_n=n!D_{[\mu_1}{}^{\mu_1} \phi\cdots D_{\mu_n]}{}^{\mu_n} \phi$,
 $$
 {\cal G}_\mn=G_\mn +2\Omega_\mn-2 \Omega_\alpha{}^\alpha g_\mn, \qquad {\cal R}=R+6 \Omega_\alpha{}^\alpha.
 $$
Ricci gauging corresponds to the case where we identify a subset of (\ref{weyl-sing}) where the $W_{\mu}$ contributions can be identified with curvature. This is only possible when $W_{\mu}$ only enters the action through $\Omega_{\mn}$, up to a total derivative \cite{gauging}. This can be guaranteed by treating $W_{\mu}$ and $\Omega_{\mu\nu}$ as independent fields and ensuring that the variation of $S_\text{weyl}$ with respect $W_{\mu}$ takes the form
\be
\frac{\delta S_\text{weyl}}{\delta W_\mu}\Big |_{\Omega \text{ fixed}}=\nabla_\nu \lambda^{\mu\nu}-2W_\nu \lambda^{\mu\nu}+\lambda^\nu{}_\nu W^\mu
\ee
where $\lambda^{\mu\nu}$ is symmetric in $\mu\nu$. This condition reduces the action to
\be \label{ricci}
S_\text{weyl}[\phi, g, \Omega] = \int d^4 x \sqrt{-g} \left[c_{2}\phi^{4} - \frac{c_{3}}{2}\phi\bar  { \cal E}_1 +c_{4}\phi^{2}{\cal R}\right]
\ee
where $c_{i}$ are dimensionless constants and the method of Ricci gauging allows one to trade $\Omega_\mn$ for $-\frac{1}{2} \left(R_\mn-\frac{1}{6} R g_\mn\right)$ as they transform identically under Weyl transformations. In general, the metric associated with these curvature terms need not be the same metric appearing in $S_\text{weyl}$ as long as it transforms the same way under Weyl transformations, hence introducing the possibility of generating a locally scale invariant bimetric, scalar theory. Initially, let's assume they are identical in which case it is comforting to note that ${\cal R}_{\mu\nu\alpha\beta}$ reduces to the Weyl tensor and consequently ${\cal{G_{\mu\nu}}} = {\cal{R}} = {0}$. We find that the resulting Ricci gauged action is
\begin{equation}
S_\text{weyl}[\phi, g] = \int d^4 x \sqrt{-g} \left[c_{2}\phi^{4} + \frac{c_{3}}{2} (\nabla \phi)^2 +\frac{c_{3}}{12}\phi^{2}R\right]
\end{equation}
and is equivalent to (\ref{local}) for appropriate choices of $c_{i}$, confirming that this action is indeed the most general subset of Horndeski possessing local scale invariance.

As briefly mentioned above, Ricci gauging opens up the possibility of generating a locally Weyl invariant bimetric, scalar theory built out of two metrics $g_{\mu\nu}$ and $h_{\mu\nu}$. Following the same procedure we find that this action takes the form
\begin{multline} \label{big}
S_\text{weyl}[\phi, g, h] = \int d^4 x \sqrt{-g} \left[c_{2}\phi^{4} + \frac{c_{3}}{2} (\nabla \phi)^2 +c_{4}\phi^{2}R 
\right.\\ 
\left.
+ \left(6c_{4} - \frac{c_{3}}{2}\right)\phi^{2}\left(\frac{1}{12}g^{\mu\nu}h_{\mu\nu}\tilde{R} - \frac{1}{2}g^{\mu\nu}\tilde{R}_{\mu\nu}\right)\right]
\end{multline} 
where $\tilde{R}$ and $\tilde{R}_{\mu\nu}$ are curvatures associated with $h_{\mu\nu}$. Because of the kinetic mixing between $g$ and $h$, the action (\ref{big}) does not fall into the class of bigravity actions presented in \cite{Hassan} , so we cannot be certain that it is ghost-free. In fact, in the light of  \cite{matas}, there are strong hints that this theory will indeed contain pathologies.

\section{Bi-scalar theories}
We now examine the case of a bi-scalar theory coupled to gravity and generalise the methods used in the previous section. Our methods made use of knowledge of the most general actions built from either $N=0$ scalars (Einstein Hilbert) or $N=1$ scalars (Horndeski), that don't exhibit the scale symmetry, to construct the most general action for $N=1$ scalars that does possess scale invariance. Therefore, to generalise these arguments and find the most general scale invariant theory of $N$ scalars coupled to gravity we require the most general theories without this symmetry for either $N-1$ or $N$ scalars. Concentrating on the case of $N=2$, let's denote our locally scale invariant action by $S_{\text{local}}[\pi, \phi, g]$. This action is unchanged by a scale tranformation $g_\mn \to \lambda^2 g_\mn, ~\pi\to \pi/\lambda, ~\phi \to \phi/\lambda$ and by choosing $\lambda(x)=\pi(x)$ we find that $S_{\text{local}}[\pi, \phi\, g] = S_{\text{local}}[1,\tilde{\phi},\tilde{g}]$, where $\tilde{\phi}=\phi/\pi$ and $\tilde{g}_{\mu\nu}=\pi^{2}g_{\mu\nu}$. As already discussed, the most general action one can construct from $\tilde{\phi}$ and $\tilde{g}_{\mu\nu}$ is the Horndeski action (\ref{horny}). Therefore the most general scale invariant theory built from two scalars and a metric with second order field equations is given by
\begin{multline}\label{2N}
S_{\text{local}}[\tilde{\phi}, \tilde{g}]=\int d^4 x \sqrt{-\tilde{g}} \left[K(\tilde{\phi}, \tilde{X})-G_3(\tilde{\phi}, \tilde{X}) {\cal \tilde{E}}_1
\right.
\\ \left.
+G_4 (\tilde{\phi}, \tilde{X})\tilde{R}+G_{4, X} {\cal \tilde{E}}_2
\right.
\\ \left. 
+G_5(\tilde{\phi}, \tilde{X}) \tilde{G}_{\mu\nu}\tilde{\nabla}^\mu \tilde{\nabla}^\nu \tilde{\phi}-\frac{G_{5, X}}{6} {\cal \tilde{E}}_3 \right].
 \end{multline}
Using the definitions of $\tilde{\phi}$ and $\tilde{g}_{\mu\nu}$, we can express this action explicity in terms of $\phi$, $\pi$ and $g_{\mu\nu}$. The relevant terms are
\ba
&&\sqrt{-\tilde{g}}=\sqrt{-g}\pi^{4}\nonumber\\
&&\tilde{R}= \pi^{-2} R-6\pi^{-3}\square \pi  \nonumber\\
&&\tilde{G}^{\mu\nu}=  \pi^{-4} G^{\mu\nu}+4\pi^{-6}\nabla^{\mu}\pi\nabla^{\nu}\pi-\pi^{-6}g^{\mu\nu}\nabla^{\kappa}\pi\nabla_{\kappa}\pi\nonumber\\
&&\qquad \qquad-2\pi^{-5}\nabla^{\mu}\nabla^{\nu}\pi+2g^{\mu\nu}\pi^{-5}\square\pi \nonumber\\
&&\tilde{X}=\pi^{-4} X_{\phi\phi}-2\phi\,\pi^{-5}X_{\phi\pi}+\phi^{2}\,\pi^{-6}X_{\pi\pi} \nonumber\\
&&\tilde{\nabla}_{\mu}\tilde{\nabla}_{\nu}\tilde{\phi}= \pi^{-1}\nabla_{\mu}\nabla_{\nu}\phi-\phi\,\pi^{-2}\nabla_{\mu}\nabla_{\nu}\pi-4\pi^{-2}\nabla_{(\mu}\phi\nabla_{\nu)}\pi\nonumber\\
&&\qquad \qquad +\pi^{-2} g_{\mu\nu}\nabla^{\alpha}\pi\nabla_{\alpha}\phi+4\phi\,\pi^{-3}\nabla_{\mu}\pi\nabla_{\nu}\pi\nonumber\\
&&\qquad \qquad-\phi\,\pi^{-3}g_{\mu\nu}\nabla^{\alpha}\pi\nabla_{\alpha}\pi \nonumber\\
&&\tilde{{\cal E}}_1=\tilde{\square}\tilde{\phi}=\pi^{-3}\square\phi-\phi\,\pi^{-4}\square\pi \nonumber
\ea
\ba
&&\tilde{{\cal E}}_2=2\delta^{\mu_{1}}_{[\mu_{2}}\delta^{\mu_{3}}_{\mu_{4}]}\big(
\pi^{-3}\nabla_{\mu_{1}}\nabla^{\mu_{2}}\phi-\phi\pi^{-4}\nabla_{\mu_{1}}\nabla^{\mu_{2}}\pi\nonumber\\
&&-2\pi^{-4}\nabla_{\mu_{1}}\phi\nabla^{\mu_{2}}\pi-2\pi^{-4}\nabla^{\mu_{2}}\phi\nabla_{\mu_{1}}\pi+\pi^{-4} \delta_{\mu_{1}}^{\mu_{2}}\nabla^{\alpha}\pi\nabla_{\alpha}\phi\nonumber\\
&& +4\phi\,\pi^{-5}\nabla_{\mu_{1}}\pi\nabla^{\mu_{2}}\pi-\phi\,\pi^{-5}\delta_{\mu_{1}}^{\mu_{2}}\nabla^{\alpha}\pi\nabla_{\alpha}\pi\big)\nonumber\\
&& \big(
\pi^{-3}\nabla_{\mu_{3}}\nabla^{\mu_{4}}\phi-\phi\,\pi^{-4}\nabla_{\mu_{3}}\nabla^{\mu_{4}}\pi-2\pi^{-4}\nabla_{\mu_{3}}\phi\nabla^{\mu_{4}}\pi\nonumber\\
&&-2\pi^{-4}\nabla^{\mu_{4}}\phi\nabla_{\mu_{3}}\pi+\pi^{-4} \delta_{\mu_{3}}^{\mu_{4}}\nabla^{\alpha}\pi\nabla_{\alpha}\phi\nonumber\\
&& +4\phi\,\pi^{-5}\nabla_{\mu_{3}}\pi\nabla^{\mu_{4}}\pi-\phi\,\pi^{-5}\delta_{\mu_{3}}^{\mu_{4}}\nabla^{\alpha}\pi\nabla_{\alpha}\pi\big)\nonumber
\ea
\ba
&&\tilde{{\cal E}}_3=6\delta^{\mu_{1}}_{[\mu_{2}}\delta^{\mu_{3}}_{\mu_{4}}\delta^{\mu_{5}}_{\mu_{6}]}\big(
\pi^{-3}\nabla_{\mu_{1}}\nabla^{\mu_{2}}\phi-\phi\,\pi^{-4}\nabla_{\mu_{1}}\nabla^{\mu_{2}}\pi\nonumber\\
&&-2\pi^{-4}\nabla_{\mu_{1}}\phi\nabla^{\mu_{2}}\pi-2\pi^{-4}\nabla^{\mu_{2}}\phi\nabla_{\mu_{1}}\pi+\pi^{-4} \delta_{\mu_{1}}^{\mu_{2}}\nabla^{\alpha}\pi\nabla_{\alpha}\phi\nonumber\\
&& +4\phi\,\pi^{-5}\nabla_{\mu_{1}}\pi\nabla^{\mu_{2}}\pi-\phi\,\pi^{-5}\delta_{\mu_{1}}^{\mu_{2}}\nabla^{\alpha}\pi\nabla_{\alpha}\pi\big)\nonumber\\
&& \big(
\pi^{-3}\nabla_{\mu_{3}}\nabla^{\mu_{4}}\phi-\phi\,\pi^{-4}\nabla_{\mu_{3}}\nabla^{\mu_{4}}\pi-2\pi^{-4}\nabla_{\mu_{3}}\phi\nabla^{\mu_{4}}\pi\nonumber\\
&&-2\pi^{-4}\nabla^{\mu_{4}}\phi\nabla_{\mu_{3}}\pi+\pi^{-4} \delta_{\mu_{3}}^{\mu_{4}}\nabla^{\alpha}\pi\nabla_{\alpha}\phi\nonumber\\
&& +4\phi\,\pi^{-5}\nabla_{\mu_{3}}\pi\nabla^{\mu_{4}}\pi-\phi\,\pi^{-5}\delta_{\mu_{3}}^{\mu_{4}}\nabla^{\alpha}\pi\nabla_{\alpha}\pi\big)\nonumber\\
&& \big(
\pi^{-3}\nabla_{\mu_{5}}\nabla^{\mu_{6}}\phi-\phi\,\pi^{-4}\nabla_{\mu_{5}}\nabla^{\mu_{6}}\pi-2\pi^{-4}\nabla_{\mu_{5}}\phi\nabla^{\mu_{6}}\pi\nonumber\\
&&-2\pi^{-4}\nabla^{\mu_{6}}\phi\nabla_{\mu_{5}}\pi+\pi^{-4} \delta_{\mu_{5}}^{\mu_{6}}\nabla^{\alpha}\pi\nabla_{\alpha}\phi\nonumber\\
&& +4\phi\,\pi^{-5}\nabla_{\mu_{5}}\pi\nabla^{\mu_{6}}\pi-\phi\,\pi^{-5}\delta_{\mu_{5}}^{\mu_{6}}\nabla^{\alpha}\pi\nabla_{\alpha}\pi\big)\nonumber
\ea
where $X_{\phi \pi} = -\frac{1}{2}\nabla \phi \nabla \pi$, for example. Note that this action is invariant under interchange of $\pi$ and $\phi$. This follows automatically from the freedom to choose the gauge parameter $\lambda$ to be either $\pi$ or $\phi$ when generating the action.

There are many models discussed in the literature which are a subset of this more general model including the scale invariant completions of the Bezrukov-Shaposhnikov actions \cite{Shap} presented in \cite{Bars2} where one of the scalars is taken to be the Higgs and the Kallosh-Linde inflationary model \cite{Linde} which we recover for $4K = -(1+ \phi^{4}/\pi^{4} - 2\phi^{2}/\pi^{2})$, $2G_{3} = -\phi/\pi$, $12G_{4} = 1-\phi^{2}/\pi^{2}$ and $G_{5} =0$. 

Again, we also compare our general model to the one presented in \cite{Bars2} and find that constraining (\ref{2N}) to have at most two derivatives in the action, reduces our bi scalar theory to the one presented there. However, in contrast to the single scalar case discussed above, we have found that many more terms can be included, which involve more than two derivatives in the action, while still keeping second order field equations. 

Following the previous section, we could also use Weyl and Ricci gauging and attempt to find this most general scale invariant action. To use this method, and keep generality, we require the corresponding $N=2$ theory without the symmetry which is, unfortunately, absent from the literature. We note that a naive generalisation of Horndeski for $N$ scalars coupled to gravity was presented in \cite{Vish}. Although it is proven to give the most general $N$ scalar action (with second order field equations) in the absence of gravity \cite{vish2}, its covariant analogue was recently shown to be missing certain terms \cite{norihiro}.  In any event, even if the most general theory were available, it is clear that the process of Ricci gauging the globally scale invariant subset would  only reproduce (\ref{2N}), a subset, or else include higher equations of motion. One could easily imagine a scenario where the later possibility is realised because the process of Ricci gauging requires one to replace first derivatives of the Weyl vector with second derivatives of the metric.

\section{N $>$ 2 scalar theories} 
We now turn our attention to the case of $N>2$ scalar fields, coupled to gravity. As mentioned, the most general covariant  action without the scaling symmetry for $N$ scalars coupled to gravity is absent from the literature so for $N>2$ scalars one cannot find the most general action with scale invariance. If and when the most general multi-scalar tensor theory is identified, let us outline how our first method should be   applied to that theory.

We denote the most general second order action of $N$ scalars coupled to gravity as $S_{N}[\phi_1, \ldots \phi_N, g]$ and are interested in scale invariant theories for which  $g_{\mu\nu} \to \lambda^2 g_\mn, ~\phi_i \to \phi_i/\lambda$, where $\lambda= \lambda(x)$. Note that we have taken each scalar to have scaling dimension $-1$, without loss of generality.  Actually, it is more convenient to further redefine our scalars, introducing $\pi_\alpha= \phi_\alpha/ \phi_N$ for $\alpha=1 \ldots (N-1)$, and $\pi_N=\phi_N$ , so that under our scale transformation $\pi_\alpha$ are invariant, and $\pi_N\to \pi_N/\lambda$. We now  write $S_{N}[\phi_1, \ldots \phi_N, g]=\bar S_{N}[\pi_1, \ldots \pi_N, g] $.

For $N$ scalars, we now let $S_\text{local}[\phi_1,  \ldots, \phi_N, g]$ denote the locally scale invariant action. Recall that this is unchanged under $g_{\mu\nu} \to \lambda^2 g_\mn, ~\phi_i \to \phi_i/\lambda$, where $\lambda=\lambda(x)$. We can now choose $\lambda=\phi_N$ as the transformation parameter and  infer $S_\text{local}[\phi_1, \ldots, \phi_N, g]=S_\text{local}[\pi_1, \ldots, \pi_{N-1},1, \tilde g]$, where the $\pi$'s are as defined in the previous paragraph, and $\tilde g_\mn=\phi_N^2 g_\mn$. We have now reduced the action to one involving a metric $\tilde g$ and only $N-1$ scalars, therefore the scale invariant subset could be easily obtained from the most general multi Horndeski action. We conclude that the {\it unique} subset of $N$-scalar Horndeski with local scale invariance is given by  $S_\text{local}[\phi_1,  \ldots, \phi_N, g]=\bar S_{N-1} [\pi_1, \ldots, \pi_{N-1}, \tilde g]$, or more explicitly
\be \label{local-multi}
S_\text{local}[\phi_1,  \ldots, \phi_N, g]=\bar S_{N-1} \left[\frac{\phi_1}{\phi_N}, \ldots, \frac{\phi_{N-1}}{\phi_N}, \phi_N^2 g\right ]
\ee
where $\bar S_{N-1}$ has the multi-Horndeski form, but with $N-1$, rather than $N$ scalars.

Although we do not yet know the form of $S_N$, we do know of a very large subset \cite{Vish}, so let us conclude this section by applying the method of Weyl and Ricci gauging to that theory, which  in four dimensions is given by \cite{Vish}
\begin{widetext}\scriptsize
\begin{multline} \label{mhorn}
\bar S_\text{N}[\pi_l, g] =\int d^4 x \sqrt{-g}\left[~A(X_{ij}, \pi_l) +  A^k(X_{ij}, \pi_l) {\cal E}_k + \frac{\partial B_{2}(X_{ij},\pi_{l})}{\partial X_{k_{1}k_{2}}}{\cal E}_{k_{1}k_{2}}
\right.
\\
\left. + \frac{1}{6}\frac{\partial B_{3}^{k_{3}}(X_{ij},\pi_l)}{\partial X_{k_{1}k_{2}}}{\cal E}_{k_{1}k_{2}k_{3}} + B_{2}(X_{ij},\pi_l)R - B_{3}^{k_{1}}(X_{ij},\pi_l)\nabla^{\mu}\nabla^{\nu}\pi_{k_{1}}G_{\mu\nu}\right]
\end{multline}
\end{widetext}
where \textit{latin} indices label the internal index of the field, $X_{ij}=-\frac{1}{2} \nabla_\mu \pi_{i} \nabla^\mu \pi_j$  for $i, j=1 \ldots N$, and
$$
{\cal E}_{k_1 \ldots k_m}=m!\nabla_{ \mu_{1}}\nabla ^{[\mu_{1}} \pi_{k_{1}} \cdots  \nabla_{\mu_{m}}\nabla^{\mu_{m}]} \pi_{k_{m}}.
$$
Recall in passing that the flat space limit of these theories are now {\it proven} to correspond to the most general multi-scalar second order theories \cite{vish2}.  

Following the same methods used for the single scalar case, we construct the following invariants:
$$
\tilde g_\mn=\pi_{N}^2 g_\mn, ~Y_{NN}=\frac{X_{NN}}{\pi_N^4},~Y_{\alpha N} =\frac{X_{\alpha N}}{\pi_N^3}, ~Y_{\alpha\beta}=\frac{X_{\alpha\beta}}{\pi_N^2}
$$
where $\alpha, \beta=1 \ldots (N-1)$. 
As expected, global scale invariance again corresponds to nothing more than dimensional analysis. 
We now Ricci gauge the global symmetry of (\ref{mhorn}). As explained in the previous sections, we introduce a Weyl vector $W_\mu$ and make replacements similar to those given in (\ref{repl}). For the multiscalar case, these are 
\ba
&&\del_\mu \pi_\alpha  \to \del_\mu \pi_\alpha\nonumber\\
&& \del_\mu \pi_N \to  D_\mu\pi_N=  \del_\mu \pi_N +W_\mu \pi_N \nonumber \\
&&\nabla_\mu\nabla_\nu \pi_\alpha \to \nabla_\mu\nabla_\nu \pi_\alpha+[2W_{(\mu} \del_{\nu)}-(W \cdot \del) g_\mn] \pi_\alpha \nonumber\\
&&\nabla_\mu \nabla_\nu \pi_N \to  D_\mn \pi_N=\nabla_\mu \nabla_\nu \pi_N +\Omega_\mn (W)\pi_N  \nonumber\\
&&\qquad \qquad +(4W_{(\mu} \delta^\alpha_{\nu)}-g_\mn W^{\alpha}) \left(\del_{\alpha}\pi_N+\frac{1}{2} W_\alpha \pi_N\right) \nonumber \\
&&R_{\mu\nu}{}^{\alpha \beta} \to {\cal R}_{\mu\nu}{}^{\alpha \beta} =R_{\mu\nu}{}^{\alpha \beta} +4\delta^{[\alpha}_{[\mu} \Omega_{\nu]}{}^{\beta]}  
\ea
where we recall that $\pi_\alpha=\phi_\alpha/\phi_N$, for $\alpha=1, \ldots (N-1)$. Note that even though these are Weyl singlets,  we still need to explictly gauge terms of the 
$\nabla_\mu\nabla_\nu \pi_\alpha$ as the metric connection is not a Weyl singlet. Again we demand that the dependence on $W_{\mu}$ only appears in the combination $\Omega_{\mu\nu}$ then Ricci gauge the remaining action. We find that the Ricci gauged version of (\ref{mhorn}) is given by
\begin{multline} \label{Ricmhorn}
S = \int d^{4}x \sqrt{-g}\left[f(Y_{\alpha\beta},\pi_{\gamma})\pi_{N}^{4}+ h(Y_{\alpha\beta},\pi_{\gamma})\pi_{N}\Box \pi_{N}
\right.
\\ \left. 
+ g^{\alpha}(Y_{\alpha\beta},\pi_{\gamma})(\pi_{N}\nabla_{\mu}\pi_{\alpha}\nabla^{\mu}\pi_{N}+ \frac{1}{2}\pi_{N}^{2} \Box \pi_{\alpha})
\right.
\\ \left. 
-\frac{1}{6}h(Y_{\alpha\beta},\pi_{\gamma})\pi_{N}^{2}R\right] 
\end{multline}
where the indices $\alpha$, $\beta$ and $\gamma$ are summed over from 1 to $N-1$ and label the index of the fields which don't transform. This action will, however, only retain second order field equations when the function $h$ is independent of $Y_{\alpha \beta}$. Clearly for the case of $N=2$ the resulting action is much less general than (\ref{2N}), highlighting that Ricci gauging is  unable to produce the most general theories. As expected, this action again reduces to the one presented in \cite{Bars2} if we ignore terms with greater than two derivatives. 

Of course, the action (\ref{mhorn}) does not include a term of the form \cite{norihiro}
\be \label{nori}
\sqrt{-g}\delta^{ik} \delta ^{jl} P^{\mu\nu \epsilon \eta} \del_\mu \pi_i \del_\nu \pi_j \del_\epsilon \pi_k \del_\eta \pi_l
\ee
or even more generally, 
\be \label{nori2}
\sqrt{-g}F^{ijkl}(\pi_m) P^{\mu\nu \epsilon \eta} \del_\mu \pi_i \del_\nu \pi_j \del_\epsilon \pi_k \del_\eta \pi_l
\ee
which nevertheless yields second order field equations. Here $P^{\mu\nu\epsilon\eta}$ is the double-dual of the Riemann tensor and $F^{ijkl}$ is symmetric on $ik$ and $jl$.  Given how the metric and the $\pi$'s transform, the globally scale invariant subsets of this are given by
\begin{multline}
\sqrt{-g}P^{\mu\nu \epsilon \eta} \left[\pi_N^{-4} f^{\alpha \rho} (\pi_\gamma) \del_\mu \pi_\alpha \del_\nu \pi_N\del_\epsilon \pi_\rho\del_\eta \pi_N \right. \\
\left.
+\pi_N^{-2} f^{\alpha \beta \rho \sigma } (\pi_\gamma)\del_\mu \pi_\alpha \del_\nu \pi_\beta \del_\epsilon \pi_\rho \del_\eta \pi_\sigma 
\right]
\end{multline}
where $f^{\alpha\rho}$ and $f^{\alpha \beta \rho \sigma }$ are symmetric in $\alpha\rho$ and $\beta\sigma$. One can now apply the method of Ricci gauging to this, but doing so will result in a theory that inevitably contains higher order field equations.

\section{Conclusions}
In this paper we have presented generalised scale invariant theories involving scalar fields coupled to gravity. Indeed, given that one of our starting points has been Horndeski's panoptic theory \cite{Horn}, we can say that (\ref{local}) and (\ref{2N}) are the most general (second order) theories of this type admitting local scale invariance for a single scalar and a bi-scalar coupled to gravity, respectively. Indeed, our proof confirms that for a single scalar, the standard conformal coupling to gravity is the {\it unique} theory with local scale invariance. When generalised to two scalar fields, however, we see a much wider range of possibilities not previously discussed in the literature. Given the wealth of applications of scale invariance, some of which we discussed in the introduction, these newly identified theories may well open up some exciting new research directions. 

We have also discussed how one would construct the most general scale invariant theory for any number of scalar fields as long as the corresponding theory without the symmetry is known. Without this knowledge, we have found a very large sub class of these theories using the technique of Ricci gauging on the globally scale invariant subset of $N$-scalar tensor theories presented in \cite{Vish}.  
Even without full generality for more than two scalars, the large subclass of scale invariant  theories  that we have now  found open up new research opportunities especially when considering multi scalar inflationary theories coming from a parent scale invariant theory.

Overall we found the technique of Weyl and Ricci  gauging to be less general than the more efficient one presented at the beginning of each section. This is not really surprising given that Ricci gauging requires the $W_{\mu}$ dependence to take a particular form. Furthermore, it will also generically generate higher order field equations, as seen in (\ref{Ricmhorn}), because Ricci gauging involves replacing first derivative pieces of the Weyl vector for second derivatives of the metric. We also note that our efficient method is more general than the one presented in \cite{faci} which is unable to construct the zeroth-order terms appearing in (\ref{2N}). 
Finally, we remark that it would be interesting to construct and study 
these scale invariant theories within the language of tractor calculus \cite{waldron}.

{\bf Acknowledgments}:~~We thank Paul Saffin, Ricardo Troncoso, Thomas Sotiriou, Eleftherios Papantonopoulos and Jorge Zanelli for  discussions.
MT thanks the School of Physics and Astronomy, Univ. of Nottingham for hospitality in the course of this work.
AP was funded by a Royal Society URF. DS was funded by an STFC studentship. MT
was funded by the FONDECYT Grant No. 3120143. The Centro de Estudios Cientificos (CECs) is funded by the Chilean Government through the Centers of Excellence Base Financing Program of Conicyt.


\begin{thebibliography}{99}


\bibitem{Planck}
  P.~A.~R.~Ade {\it et al.}  [Planck Collaboration],
  arXiv:1303.5082 [astro-ph.CO].

\bibitem{Weyl}
  H.~Weyl,
  Z.\ Phys.\  {\bf 56} (1929) 330
   [Surveys High Energ.\ Phys.\  {\bf 5} (1986) 261].

\bibitem{thooft}
  G.~'t Hooft,
  arXiv:1009.0669 [gr-qc].
  
\bibitem{Bars1}
  I.~Bars, S.~-H.~Chen and N.~Turok,
  Phys.\ Rev.\ D {\bf 84} (2011) 083513
  [arXiv:1105.3606 [hep-th]].
  I.~Bars, S.~-H.~Chen, P.~J.~Steinhardt and N.~Turok,
  Phys.\ Lett.\ B {\bf 715} (2012) 278
  [arXiv:1112.2470 [hep-th]].
  I.~Bars, S.~-H.~Chen, P.~J.~Steinhardt and N.~Turok,
  Phys.\ Rev.\ D {\bf 86} (2012) 083542
  [arXiv:1207.1940 [hep-th]].
  I.~Bars,
  arXiv:1209.1068 [hep-th].
  
\bibitem{Bars2}
  I.~Bars, P.~Steinhardt and N.~Turok,
  arXiv:1307.1848 [hep-th].
  
\bibitem{Kallosh}
  J.~J.~M.~Carrasco, W.~Chemissany and R.~Kallosh,
  arXiv:1311.3671 [hep-th].

\bibitem{Bars3}
  I.~Bars, P.~Steinhardt and N.~Turok,
  arXiv:1312.0739 [hep-th].

\bibitem{Linde}
  R.~Kallosh and A.~Linde,
  JCAP {\bf 1307} (2013) 002
  [arXiv:1306.5220 [hep-th]].

\bibitem{def}
  B.~Bellazzini, C.~Csaki, J.~Hubisz, J.~Serra and J.~Terning,
  arXiv:1305.3919 [hep-th].
  F.~Coradeschi, P.~Lodone, D.~Pappadopulo, R.~Rattazzi and L.~Vitale,
  arXiv:1306.4601 [hep-th].
  
\bibitem{seq}
  N.~Kaloper and A.~Padilla,
  arXiv:1309.6562 [hep-th].

\bibitem{Nic}
  K.~A.~Meissner and H.~Nicolai,
  Phys.\ Lett.\ B {\bf 648} (2007) 312
  [hep-th/0612165].

\bibitem{Shap}
  M.~Shaposhnikov,
  arXiv:0708.3550 [hep-th].
  F.~L.~Bezrukov and M.~Shaposhnikov,
  Phys.\ Lett.\ B {\bf 659} (2008) 703
  [arXiv:0710.3755 [hep-th]].
  M.~Shaposhnikov and D.~Zenhausern,
  Phys.\ Lett.\ B {\bf 671} (2009) 187
  [arXiv:0809.3395 [hep-th]].
  M.~Shaposhnikov and D.~Zenhausern,
  Phys.\ Lett.\ B {\bf 671} (2009) 162
  [arXiv:0809.3406 [hep-th]].
  D.~Blas, M.~Shaposhnikov and D.~Zenhausern,
  Phys.\ Rev.\ D {\bf 84} (2011) 044001
  [arXiv:1104.1392 [hep-th]].
  


\bibitem{foot}
  R.~Foot, A.~Kobakhidze and R.~R.~Volkas,
  Phys.\ Rev.\ D {\bf 82} (2010) 035005
  [arXiv:1006.0131 [hep-ph]].

\bibitem{jain}
  P.~Jain, S.~Mitra and N.~K.~Singh,
  JCAP {\bf 0803} (2008) 011
  [arXiv:0801.2041 [astro-ph]].

\bibitem{rajpoot}
  H.~Nishino and S.~Rajpoot,
  hep-th/0403039.

\bibitem{singh}
  N.~K.~Singh, P.~Jain, S.~Mitra and S.~Panda,
  Phys.\ Rev.\ D {\bf 84} (2011) 105037
  [arXiv:1106.1956 [hep-ph]].

\bibitem{adscft}
  J.~M.~Maldacena,
  Adv.\ Theor.\ Math.\ Phys.\  {\bf 2} (1998) 231
  [hep-th/9711200].


 \bibitem{biol}
  T.~Gisiger,  
  Biological Reviews 76.2 (2001): 161-209.
  
  \bibitem{psy}
 N.~Chater and G.~D.~Brown,  
 Cognition, 69(3) (1999): B17-B24.
  
\bibitem{Horn}
  G.~W.~Horndeski,
  Int.\ J.\ Theor.\ Phys.\  {\bf 10} (1974) 363.
  
\bibitem{Vish}
  A.~Padilla and V.~Sivanesan,
  JHEP {\bf 1304} (2013) 032
  [arXiv:1210.4026 [gr-qc]].

  
 \bibitem{ostro} 
  M. ~Ostrogradsky, 
  Memoires de l'Academie Imperiale des Science de Saint-Petersbourg, 4:385, 
1850. 


\bibitem{hawking}
  S.~W.~Hawking and T.~Hertog,
  Phys.\ Rev.\ D {\bf 65} (2002) 103515
  [hep-th/0107088].

\bibitem{cliff}
  C.~P.~Burgess,
  Ann.\ Rev.\ Nucl.\ Part.\ Sci.\  {\bf 57} (2007) 329
  [hep-th/0701053].


\bibitem{gauging}
  A.~Iorio, L.~O'Raifeartaigh, I.~Sachs and C.~Wiesendanger,
  Nucl.\ Phys.\ B {\bf 495} (1997) 433
  [hep-th/9607110].
  
\bibitem{luty}
  K.~Farnsworth, M.~A.~Luty and V.~Prelipina,
  arXiv:1309.4095 [hep-th].


\bibitem{dgsz}
  C.~Deffayet, X.~Gao, D.~A.~Steer and G.~Zahariade,
  Phys.\ Rev.\ D {\bf 84} (2011) 064039
  [arXiv:1103.3260 [hep-th]].
  
\bibitem{Lovelock}
  D.~Lovelock,
  J.\ Math.\ Phys.\  {\bf 12} (1971) 498.
  
\bibitem{deser}
  S.~Deser, Ann. Phys. {\bf{59}} 248

\bibitem{Tsouk}
R.~Troncoso, M.~Tsoukalas, CECS-PHY-11/11 

\bibitem{alvarez}
E.~Alvarez, D.~Blas, J.~Garriga and E.~Verdaguer,
  Nucl.\ Phys.\ B {\bf 756} (2006) 148
  [hep-th/0606019].


\bibitem{Hassan}
  S.~F.~Hassan and R.~A.~Rosen,
  JHEP {\bf 1202} (2012) 126
  [arXiv:1109.3515 [hep-th]].
  
\bibitem{matas}
  C.~de Rham, A.~Matas and A.~J.~Tolley,
  arXiv:1311.6485 [hep-th].

\bibitem{vish2}
  V.~Sivanesan,
  arXiv:1307.8081 [gr-qc].
  
\bibitem{norihiro}
  T.~Kobayashi, N.~Tanahashi and M.~Yamaguchi,
  Phys.\  Rev.\  D 88, {\bf 083504} (2013)
  [arXiv:1308.4798 [hep-th]].

\bibitem{faci}
  S.~Faci,
  Europhys.\ Lett.\  {\bf 101} (2013) 31002.

\bibitem{waldron}
  R.~Bonezzi, O.~Corradini and A.~Waldron,
  J.\ Phys.\ Conf.\ Ser.\  {\bf 343} (2012) 012128
  [arXiv:1003.3855 [hep-th]].


\end{thebibliography}
\end{document}